\documentclass[journal=jpclcd,manuscript=article]{achemso}

\usepackage{url}
\usepackage{amssymb}
\usepackage{amsmath}
\usepackage{float}
\usepackage{natbib}
\usepackage{subcaption}
\usepackage[hidelinks]{hyperref}

\UseRawInputEncoding

\usepackage{caption}
\DeclareCaptionLabelSeparator{fullstop}{.\ }
\usepackage{etoolbox}
\makeatletter
\patchcmd{\acs@contact@details}{E}{*\,E}{}{}
\def\acs@author@fnsymbol#1{}
\makeatother

\usepackage[pagewise]{lineno}

\author{Abdul Rehman${^\dagger{^*}}$}
\email{a.rehman@utwente.nl}

\author{Robbert W.E. van de Kruijs$^\dagger$}
\author{Wesley T.E. van den Beld$^\dagger$}
\author{Jacobus M. Sturm$^\dagger$}
\author{Marcelo Ackermann$^\dagger$}

\affiliation{$^\dagger$Industrial Focus Group XUV Optics, MESA+ Institute for Nanotechnology, University of Twente, Drienerlolaan 5, 7522NB Enschede, the Netherlands}

\title[\textsf{achemso}]
{Engineering Work Function to Stabilize Metal Oxides in Reactive Hydrogen}

\begin{document}

\begin{tocentry}
    \includegraphics[width=1.0\linewidth]{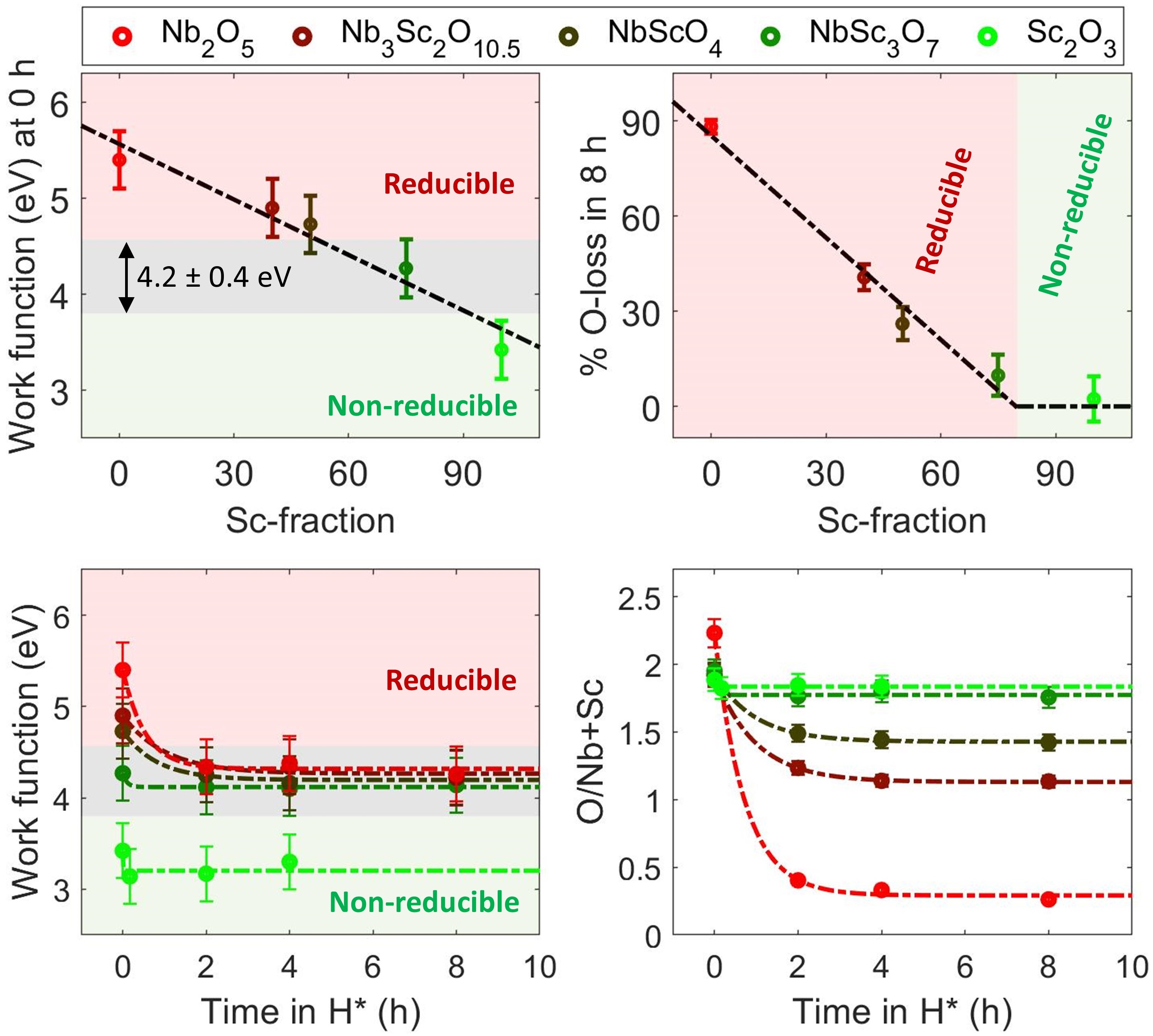}
    \centering
\end{tocentry}

\begin{abstract}
Hydrogen, crucial for the green energy transition, poses a challenge due to its tendency to degrade surrounding wall materials. To harness hydrogen's potential, it is essential to identify materials’ parameter(s) that modulate hydrogen-material interaction. In a recent publication, we have shown that the reduction (de-nitridation) of transition metal (TM)-nitrides in hydrogen radicals (H*) stops when their work function drops below a threshold limit. In this work, we tailor the work function of a complex TM-oxide by tuning the relative content of its constituent TM-atoms. We show that increasing the fraction of a low work function TM decreases the work function of the complex oxide, thereby decreasing its reducibility (de-oxidation) in H*. This leads to the stabilization of the higher oxidation states of a high work function TM, which otherwise readily reduce in H*. We propose that the work function serves as a tuneable parameter, modulating the interaction of hydrogen with TM compounds. 
\end{abstract}


Hydrogen plays a key role in a wide range of applications, from green energy solutions such as fusion \cite{C4RN1}, energy storage \cite{C4RN2, C4RN4}, and transport \cite{C4RN3} to advanced semiconductor fabrication, where it serves as an etchant \cite{C4RN5, C4RN1300}, and a reducing agent \cite{C4RN1700, C4RN7}. Nevertheless, the tendency of hydrogen to react with and diffuse into the surrounding wall materials poses a high operational risk, e.g., embitterment, blistering, interface defects, and chemical erosion \cite{C4RND1, C4RND3, C4RND2, C4RN17, C4RN23, C4RN9}. Hence, to fully realize hydrogen's potential, it is essential to develop novel coatings that are stable in reactive-hydrogen environments and can protect hydrogen-sensitive system components. The development of such coatings necessitates a strategic approach to effectively modulate the interaction between materials and hydrogen.  

In recent publications \cite{C4RN10, C4RN2233}, we demonstrate that the chemical stability of transition metal nitrides (TMNs) in high-temperature hydrogen radical (H*) environments depends on their work function. When the work function of a TMN system drops below a threshold value ($\phi_{TH}$ = 4.3 $\pm$ 0.4 eV, in H* at elevated temperatures), its reduction (de-nitridation) effectively stops - even though further reduction remains thermodynamically favorable, i.e., negative change in the Gibbs free energy ($\Delta$G) for the reduction reaction: TMN$_\mathrm{y}$ + xH $\rightarrow$ TMN$_{\mathrm{y}-{1}}$ + NH$_\mathrm{x}$. We explain this by the preferential binding of H* to transition metal (TM)-atoms \cite{C4RN12, C4RN13}, which impedes the formation of volatile NH$_\mathrm{x}$ species. 

In this work, we demonstrate that the work function serves as a tunable parameter, enabling control over the reduction (de-oxidation) of (complex) TM oxides in H*. By strategically alloying a high work function oxide with a lower work function oxide, we show that the work function of the resulting complex oxide can be modulated by tuning the relative fraction of its constituent TM-atoms. This shift in the work function correlates directly with the reduction of the complex oxide in H*, with a lower work function leading to a lower reduction. Furthermore, we show that the higher oxidation states of a high work function TM in the complex oxide are stable in H*, which otherwise reduce readily.

For our study, we selected Nb$_2$O$_5$, Sc$_2$O$_3$, and their complex oxides (NbSc$_\mathrm{y}$O$_\mathrm{x}$). Nb$_2$O$_5$ and Sc$_2$O$_3$ present extreme cases for our study. Nb$_2$O$_5$ has a high work function of $\approx$ 5.2 eV \cite{C4RN14}, which, according to our work function model \cite{C4RN10}, is expected to undergo significant reduction during H*-exposure. In contrast, Sc$_2$O$_3$ due to its lower work function ($\approx$ 3.5 eV \cite{C4RN27}) is expected to be non-reducible in H*.

Given that the formation of NbSc$_\mathrm{y}$O$_\mathrm{x}$ is energetically feasible \cite{C4RN18, C4RN19}, we can modulate its work function by altering the relative proportions of Nb- and Sc-atoms. For this study, we specifically chose NbSc$_\mathrm{y}$O$_\mathrm{x}$ compositions with approximately 40\%, 50\%, and 75\% Sc-atoms relative to Nb-atoms, aiming to vary the work function of NbSc$_\mathrm{y}$O$_\mathrm{x}$ around $\phi_{TH}$ \cite{C4RN10}. As a starting point, we estimated the work function of NbSc$_\mathrm{y}$O$_\mathrm{x}$ using a compositional weighted average of the work functions of its constituent TM-oxides \cite{C4RN30}. We recognize that this estimation does not take into account critical factors such as surface termination, orientation, and unique hetero-structuring in complex oxides, all of which influence the actual work function \cite{C4RN29}. Nevertheless, this estimation provides a baseline for approximating the trend in the work function of a complex oxide as the relative proportion of its constituent TM-atoms is varied.

We exposed 5 $\pm$ 0.5 nm thin films of the selected materials to H* at 550 $^\circ$C for a total duration of 8 h. These H*-exposure conditions are relevant for the development of hydrogen-protective coatings for EUV scanners and fusion reactors \cite{C4RN24, C4RN23, C4RN26, C4RN22}. The Angle-Resolved X-ray Photoelectron Spectroscopy (AR-XPS) measurements are performed on the samples after 2 h, 4 h, and 8 h of H*-exposure. Since the Sc$_2$O$_3$ sample is effectively non-reducible, it is exposed to H* for only 4 h, with AR-XPS measurements performed after 10 min, 2 h, and 4 h. To saturate thermally induced processes prior to high-temperature H*-exposures, the samples are annealed in a vacuum for 2 h at 550 $^\circ$C. These samples, referred to as pre-exposed (0 h) in the text and figures, are used as a reference to assess the reducibility of the oxides. The samples exposed to H* are referred to by their total H*-exposures time: 10 min, 2 h, 4 h, and 8 h H*-exposed. The stoichiometry/spectra of the samples mentioned in the text and figures correspond to the XPS measurements taken at a take-off angle ($\Theta$) of 34.25$^\circ$. Note that, work function (surface average) measurements are also performed via XPS in normal lens mode \cite{C4RN10}. 

In the subsequent paragraphs, we first discuss that as Sc-fraction in the NbSc$_\mathrm{y}$O$_\mathrm{x}$ increases, both its work function and reduction decrease. Next, we discuss that this decreased reduction leads to the stabilization of higher oxidation states of Nb, where these oxidation states in Nb$_2$O$_5$ readily reduce. 

\begin{figure}[H]
    \centering
    \includegraphics[width=0.7\linewidth]{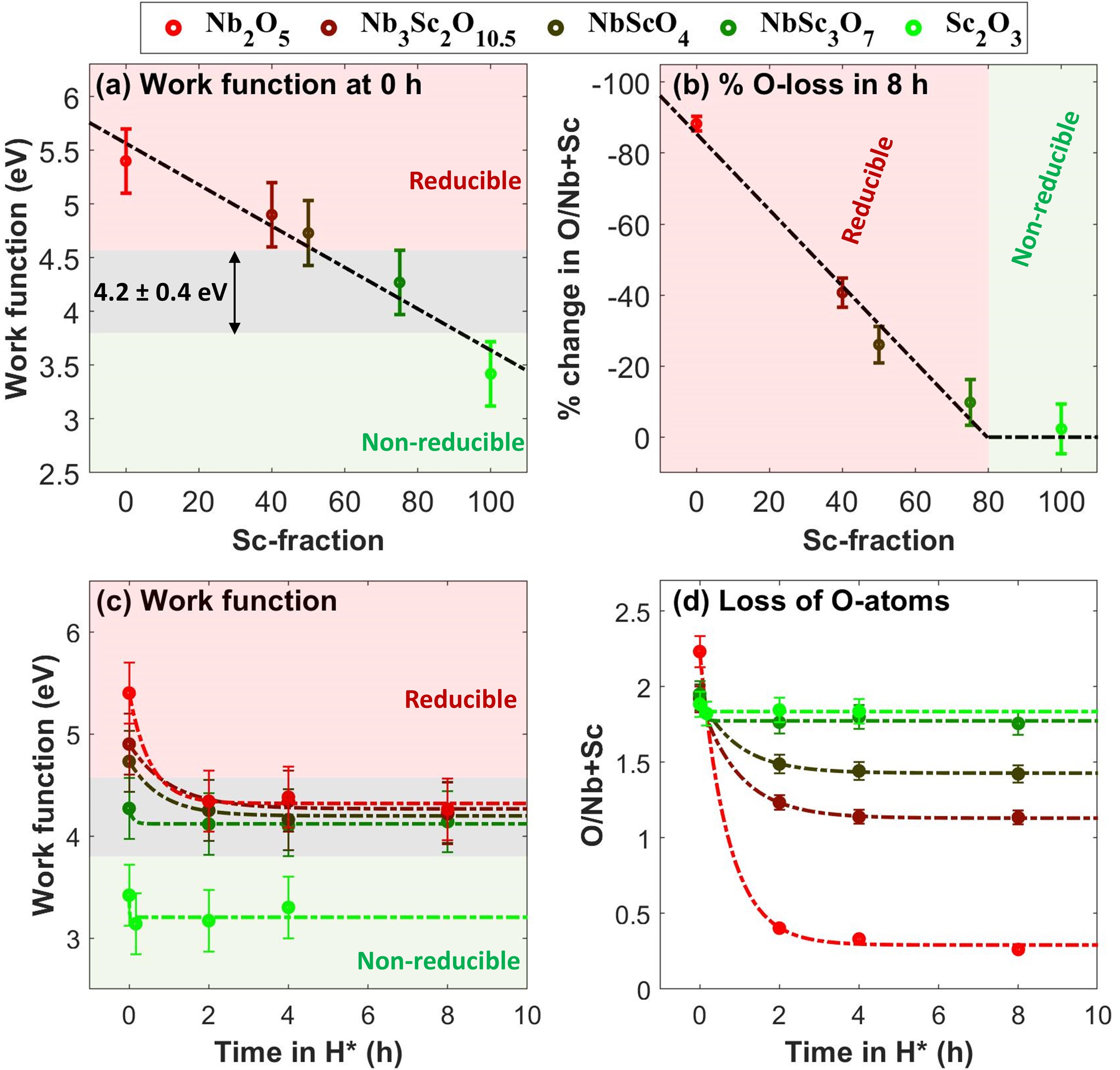}
    \caption{Measured work functions, \% O-loss, and O/Nb+Sc ratios in the Nb$_2$O$_5$, Nb$_3$Sc$_2$O$_{10.5}$, NbScO$_4$, NbSc$_3$O$_7$, and Sc$_2$O$_3$ samples. (a) The work function of the pre-exposed (0 h) samples decreases almost linearly with increasing Sc-fraction. (b) This decrease in the work function leads to a decrease in the O-loss in the samples upon H*-exposure, calculated as the \% change in the O/Nb+Sc ratio$^*$. (c-d) As O-atoms are removed from the samples during H*-exposure, their work function progressively decreases, until reaching a stable value of 4.2 $\pm$ 0.4 eV. At this threshold ($\phi_{TH}$), the reduction reaction effectively stops. \\
    * $\% \, \mathrm{O\text{-}loss} = \left( \frac{\mathrm{O/Nb+Sc \; after \; the \; last \; H^*\text{-}exp} - \mathrm{O/Nb+Sc \; before \; H^*\text{-}exp}}{\mathrm{O/Nb+Sc \; before \; H^*\text{-}exp}} \right) \times 100$}
    \label{C4_O_TM_WF}
\end{figure}
 
\begin{figure}[H]
    \centering
    \includegraphics[width=0.6\linewidth]{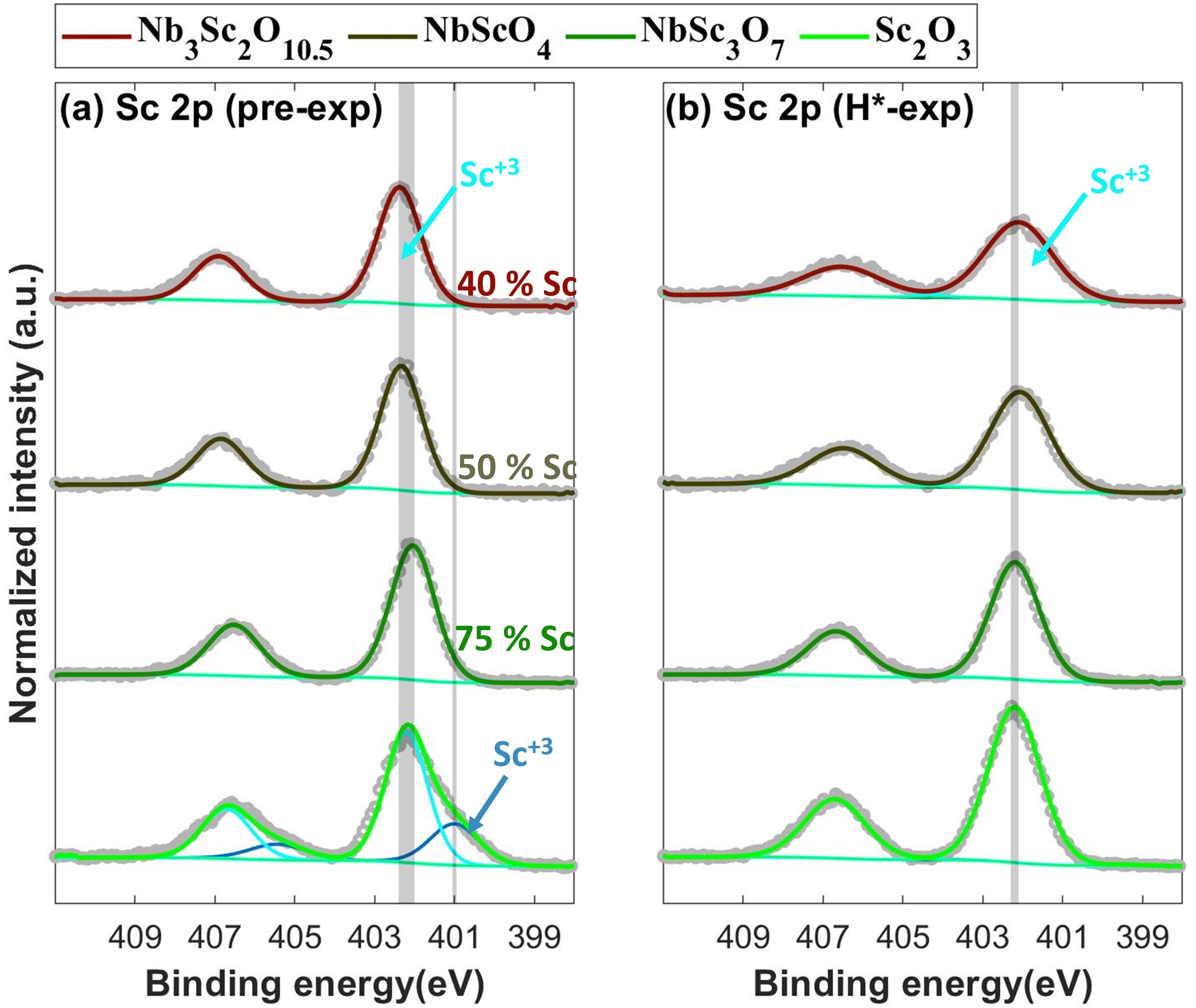}
    \caption{Sc 2p XPS spectra of (a) the pre-exposed and (b) the 8 h H*-exposed Nb$_3$Sc$_2$O$_{10.5}$, NbScO$_4$, and NbSc$_3$O$_7$,  along with 4 h H*-exposed Sc$_2$O$_3$ samples. In all the samples, Sc-atoms are in a $+3$ oxidation state, with the Sc$_2$O$_3$ doublet appearing at lower binding energy and the ScOOH doublet at higher binding energy. For the samples where O-loss is substantial (Nb$_3$Sc$_2$O$_{10.5}$ and NbScO$_4$), the Sc 2p spectra show significant changes upon H*-exposure. No significant change in the Sc 2p spectra of the NbSc$_3$O$_7$ sample is observed. The changes in the Sc 2p spectra of the Sc$_2$O$_3$ sample are attributed to the hydrogenation upon H*-exposure.}
    \label{C4_Sc2p}
\end{figure}

In line with our model \cite{C4RN10}, the Nb$_2$O$_5$ sample with a high work function, undergoes strong reduction upon H*-exposure (Figure \ref{C4_O_TM_WF}). The strong reduction of the sample is evident from the pronounced decrease in the O/Nb ratio (Figure \ref{C4_O_TM_WF}b and \ref{C4_O_TM_WF}d). Inversely, the Sc$_2$O$_3$ sample is effectively non-reducible due to its low work function, as evident by the minimal change in the O/Sc ratio upon H*-exposure (Figure \ref{C4_O_TM_WF}). Note that the O/Sc ratio in the Sc$_2$O$_3$ sample is approximately 1.9 (Figure \ref{C4_O_TM_WF}d). The high O-fraction in the sample is attributed to the formation of ScOOH (Figure \ref{C4_Sc2p}) \cite{C4RNScO1, C4RN32, C4RN33}.

The work function of the NbSc$_\mathrm{y}$O$_\mathrm{x}$ samples exhibits a clear dependence on the Sc-fraction. Notably, the work function of the pre-exposed (0 h) NbSc$_\mathrm{y}$O$_\mathrm{x}$ samples decreases almost linearly with increasing Sc-fraction, consistent with our estimation (Figure \ref{C4_O_TM_WF}a). Furthermore, the offset between the measured and linearly approximated work functions is a couple tenths of eV. This agreement can be attributed to the amorphous/nanocrystalline morphology of the samples, effectively averaging variations in surface termination, orientation, and heterostructuring (Figure S1-S3). 

The decrease in the work function with increasing Sc-fraction is correlated with a smaller drop in the O/Nb+Sc ratios upon H*-exposure (Figure \ref{C4_O_TM_WF}a and \ref{C4_O_TM_WF}b). Specifically, the change in both the work function and the O/Nb+Sc ratio upon H*-exposures is smaller for the samples with higher Sc-fraction (Figure \ref{C4_O_TM_WF}c and \ref{C4_O_TM_WF}d). This indicates that the extent of oxide reduction in H* is directly related to its work function, i.e., decreasing with decreasing work function.

Notably, the reduction reaction of all the samples effectively stops as their work function drops to 4.2 $\pm$ 0.4 eV (Figure \ref{C4_O_TM_WF}c and \ref{C4_O_TM_WF}d), aligning with our model \cite{C4RN10}. Since, a sufficient number of O-atoms remain at the surface level following the last H*-exposure (Figures S4, S5b, S6b, S7b, and S8), the reduction reaction is not limited by the diffusion of subsurface O-atoms to the surface \cite{C4RN10}. Therefore, based on the modeling by Van de Walle et al. \cite{C4RN12, C4RN13}, we propose that when the work function of an oxide system is higher than 4.2 $\pm$ 0.4 eV, H* adsorption on O-atoms is favorable, enabling the formation of volatile OH$_\mathrm{x}$ species \cite{C4RN10}. However, as electronegative atoms (in this case, O-atoms) are removed \cite{C4RN36}, or electropositive atoms (in this case, Sc-atoms) are incorporated, the work function of the oxide system decreases, eventually reaching the 4.2 $\pm$ 0.4 eV threshold limit (Figure \ref{C4_O_TM_WF}b). At this point, H* preferentially bind to TM-atoms instead of O-atoms, making the formation of OH$_\mathrm{x}$ unfavorable.

Due to the work function threshold limit, the pre-exposed samples with a lower work function exhibit a higher O/Nb+Sc ratio following the last H*-exposure (Figure \ref{C4_O_TM_WF}d), suggesting that higher oxidation states of Nb- and Sc-atoms are stabilized with increasing Sc-fraction. We discuss the Nb 3d and Sc 2p XPS spectra of the samples in the subsequent paragraphs, which provide insights into the oxidation states of Nb- and Sc-atoms in the NbSc$_\mathrm{y}$O$_\mathrm{x}$ samples.

\begin{figure}[H]
    \centering
    \includegraphics[width=0.6\linewidth]{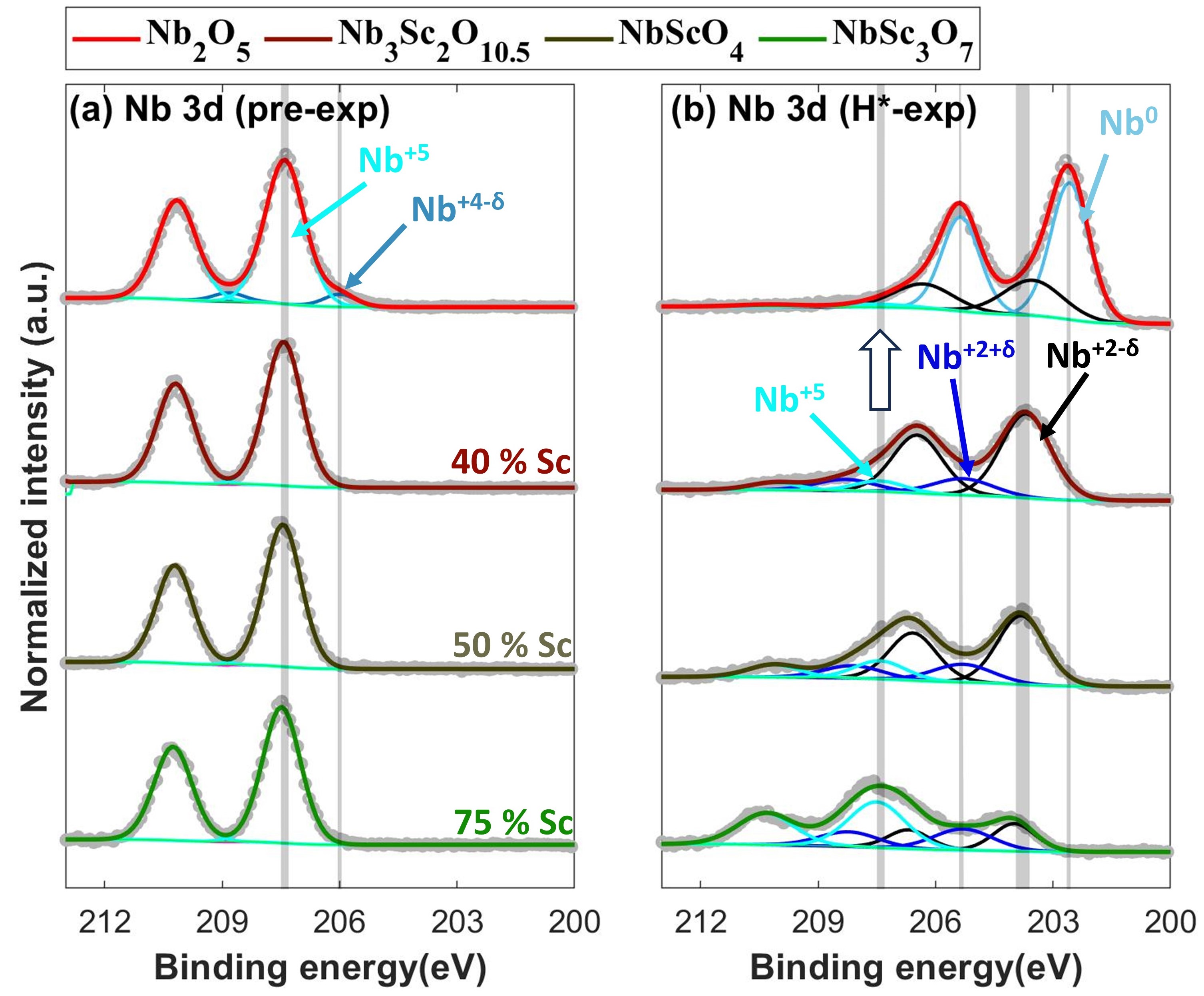}
    \caption{Nb 3d XPS spectra of (a) the pre-exposed and (b) the 8 h H*-exposed Nb$_2$O$_5$, Nb$_3$Sc$_2$O$_{10.5}$, NbScO$_4$, and NbSc$_3$O$_7$ samples. In the pre-exposed samples, Nb-atoms are predominantly in a +5 oxidation state. The oxidation states of Nb-atoms in the post-H*-exposed samples depend on the Sc-fraction in NbSc$_\mathrm{y}$O$_\mathrm{x}$. A higher Sc-fraction results in a greater proportion of Nb-atoms remaining in higher oxidation states after H*-exposure. This suggests that the samples predominantly contain NbSc$_\mathrm{y}$O$_\mathrm{x}$, rather than distinct NbO$_\mathrm{x}$ and ScO$_\mathrm{x}$ phases.}
    \label{C4_Nb3d}
\end{figure}
    
In all pre-exposed samples, Nb-atoms are in a +5 oxidation state, except in Nb$_2$O$_5$, where a minor fraction of Nb-atoms exhibit a $+4-\delta$ oxidation state (Figure \ref{C4_Nb3d}a). After 8 h H*-exposure, the oxidation states of Nb-atoms depend on the Sc-fraction (Figure \ref{C4_Nb3d}b).  The Nb-atoms in the samples with higher Sc-fraction exhibit a higher fraction of higher oxidation states after 8 h H*-exposure. For instance, in the Nb$_2$O$_5$ sample, $\approx$ 28\% of Nb-atoms are in a $+2-\delta$ oxidation state(s), with the remainder being metallic (Nb$^\circ$). In comparison, the samples with approximately 40\%, 50\%, and 75\% Sc show approximately 10\%, 17\%, and 51\% of Nb-atoms in a +5 oxidation state, respectively. The rest of the Nb-atoms in these samples are distributed among $+2\pm\delta$ oxidation states (Figure \ref{C4_Nb3d}b). The presence of multiple oxidation states of Nb-atoms in the H*-exposed samples is attributed to the under-stoichiometric O/Nb ratio, due to which O-atoms are distributed among Nb-atoms in such a way that minimizes the formation energy. The stabilization of higher oxidation states of Nb-atoms with increasing Sc-atoms fraction in NbSc$_\mathrm{y}$O$_\mathrm{x}$ suggests that the incorporation of Sc-atoms (low work function) can effectively lower the reducibility of Nb$_2$O$_5$ (higher work function) in H*. This is analogous to the decreasing reducibility of Nb$_2$O$_5$ as O-atoms are removed \cite{C4RN36} (Figure \ref{C4_O_TM_WF}c and \ref{C4_O_TM_WF}d).

Unlike the Nb 3d XPS spectra, changes in the Sc 2p XPS spectra of the samples are less pronounced (Figure \ref{C4_Sc2p}). In both the pre- and the post-H*-exposed samples, the Sc-atoms are in a $+3$ oxidation state (Figure \ref{C4_Sc2p}). This is due to a high O-fraction in the samples relative to Sc-atoms, combined with Sc's high oxidation potential. However, in the samples where substantial O-loss occurred during H*-exposure, significant changes in the Sc 2p are observed. For instance, in the samples with approximately 40\% and 50\% Sc, the Sc$^{+3}$ doublet shifted by $\approx$ 0.3 eV towards a lower binding energy upon 8 h H*-exposure. Furthermore, the full width at half maximum (FWHM) of the fitted doublet is increased by $\approx$ 57\% and $\approx$ 37\%, respectively, compared to the pre-exposed sample (Figure \ref{C4_Sc2p}). This indicates O-loss during H*-exposure. In contrast, in the sample with 75\% Sc, where the O/Nb+Sc ratio did not decrease significantly, no major change in the Sc 2p spectra is noted. The change in the Sc 2p spectra of the Sc$_2$O$_3$ sample after H*-exposure is due to the hydrogenation of the sample (Figure \ref{C4_Sc2p}) \cite{C4RNScO1}. Overall, the evolution of the Sc 2p XPS spectra with increasing Nb-fraction suggests that the incorporation of (higher work function) Nb-atoms in Sc$_2$O$_3$ (lower work function) increases its reducibility in H*.

The stabilization of Nb$^{+5}$ (Figure \ref{C4_Nb3d}) and Sc$^{+3}$ (Figure \ref{C4_Sc2p}) with increasing Sc-fraction indicate that the samples predominantly contain NbSc$_\mathrm{y}$O$_\mathrm{x}$ compound(s). If separate NbO$_\mathrm{x}$ and ScO$_\mathrm{x}$ phases were present, the Nb 3d and Sc 2p XPS spectra of the NbSc$_\mathrm{y}$O$_\mathrm{x}$ samples would be similar to those of the Nb$_2$O$_5$ and Sc$_2$O$_3$ samples, respectively. This is consistent with the low kinetic energy (LKE) XPS spectra and Kelvin Probe Atomic Force Microscopy (KPAFM) (Figure S3 and S14), which did not show the presence of phases with significantly different work functions on the surface, as would be expected for NbO$_\mathrm{x}$ and ScO$_\mathrm{x}$. Furthermore, Transmission Electron Microscopy Energy-Dispersive X-ray Spectroscopy (TEM-EDS) confirms the homogeneous distribution of TM- and O-atoms across the sample's depth (Figure S2). 

Our results demonstrate that alloying NbO$_\mathrm{x}$ and ScO$_\mathrm{x}$ predominantly results in the formation of NbSc$_\mathrm{y}$O$_\mathrm{x}$ (complex oxides), with the reduction of NbSc$_\mathrm{y}$O$_\mathrm{x}$ in H* dependent on the Nb/Sc ratio. As the Sc-fraction increases, the work function of the NbSc$_\mathrm{y}$O$_\mathrm{x}$ decreases. This decrease in the work function leads to a decrease in the reduction of NbSc$_\mathrm{y}$O$_\mathrm{x}$, thereby stabilizing higher oxidation states of Nb- and Sc-atoms. Additionally, this study shows that the reduction reaction on all the studied oxide systems stops when their work function reaches a 4.2 $\pm$ 0.4 eV threshold, consistent with our recent publication \cite{C4RN10}.

Based on our findings, we propose that the work function of a TM-compound serves as a predictive tuneable parameter, governing its chemical stability in reactive hydrogen environments. Lowering the work function effectively decreases TM-compound's reducibility. Furthermore, our work demonstrates that by modulating the work function, specific oxidation states of a TM can be stabilized in a hydrogen environment. These insights provide a valuable framework for designing TM-compound hydrogen-protective coatings by strategically engineering their work function.

\section{Methodology}\label{Methodology}
Thin films of Nb$_2$O$_5$, Nb$_3$Sc$_2$O$_{10.5}$, NbScO$_4$, NbSc$_3$O$_7$, and Sc$_2$O$_3$ are deposited onto Si(100) substrates through reactive DC magnetron co-sputtering using Nb and Sc targets. The deposition chamber maintains a base pressure of low 10$^{-8}$ mbar. Ar (99.999\%) and O$_2$ (99.999\%) with both a flow rate of 20 sccm are used as sputtering gases, providing a working pressure of 10$^{-3}$ mbar. The deposition rates of Nb$_2$O$_5$ and Sc$_2$O$_3$ were calibrated as a function of the magnetron currents. Approximately 20 nm thick films were deposited for calibrations. Film thicknesses are measured using X-ray Reflectivity (XRR), performed using a Malvern Panalytical Empyrean laboratory diffractometer, which employs a monochromatic Cu-K$\alpha$1 radiation source. The magnetron currents are then adjusted to achieve approximately 40\%, 50\%, and 75\% Sc-atoms relative to Nb-atoms in the final samples.

5 $\pm$ 0.5 nm thin films are deposited, where the thickness is controlled by the deposition time (Table S1). The thickness of the sample is chosen to ensure that the entire depth of the samples is probed by Angle-Resolved X-ray Photoelectron Spectroscopy (AR-XPS). XPS measurements are performed using a Thermo-Fisher Theta Probe Angle-Resolved X-ray Photoelectron spectrometer, which utilizes a monochromatic Al-K$\alpha$ radiation source. During AR-XPS measurements, spectra are collected at take-off angles ($\Theta$: from sample's surface normal), ranging from 26.75$^\circ$ to 71.75$^\circ$, providing probing depth ranging from $\approx$ 5 nm to $\approx$ 1.5 nm, respectively, with a spot size of 400$\times$400 $\mu$$m$. Note that for quantification, we calibrated AR-XPS signals based on the transmission function measured for the normal (not angle-resolved) lens mode. Furthermore, the stoichiometry of the samples mentioned in the text and figures is measured at $\Theta$=34.25$^\circ$, with an uncertainty of $\pm$ 10\%.

\begin{figure}[H]
    \centering
    \includegraphics[width=1.0\linewidth]{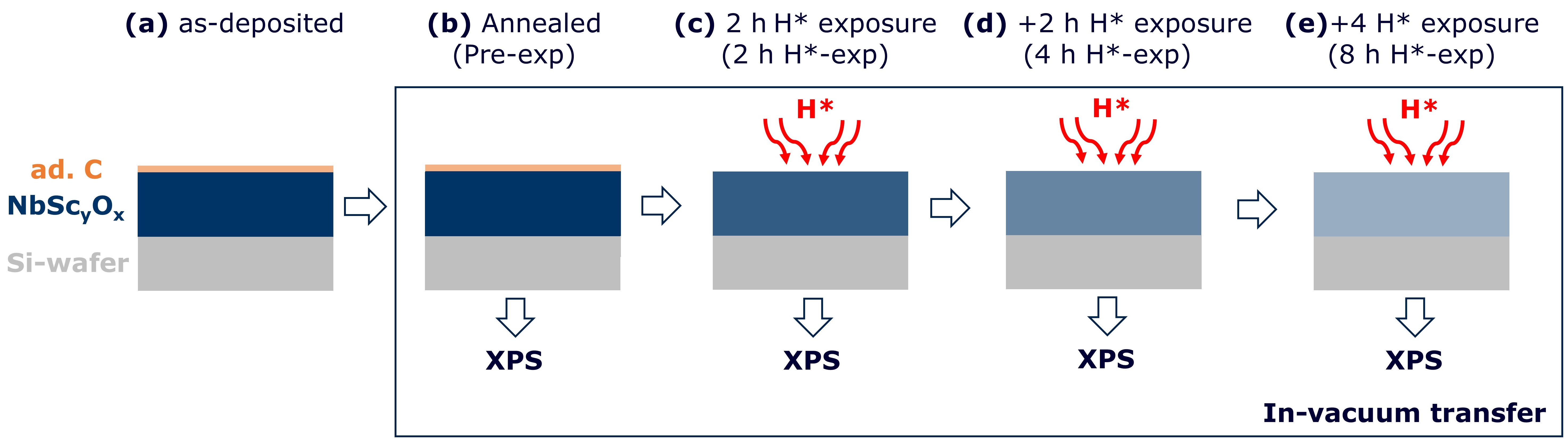}
    \caption{Schematic of methodology. (a) NbSc$_\mathrm{y}$O$_\mathrm{x}$ samples are deposited via reactive DC magnetron co-sputtering. A thin layer of adventitious carbon (ad. C) formed on the samples' surfaces during ambient storage. (b) The samples are first annealed at 550 $^\circ$C for 2 h. (c-e) The samples are then exposed to H* for a total of 8 h at 550 $^\circ$C, where XPS measurements are performed on the samples after (c) 2 h, (d) 4 h, and (e) 8 h. The samples are transferred in vacuum (lower range of 10$^{-9}$ mbar) between the processing (annealing/H*-exposure) and the XPS chambers. }
    \label{C4_methodology}
\end{figure}
 
After deposition, samples are stored in ambient for approximately a week, during which they accumulate adventitious carbon (ad. C). The presence of O-atoms in the ad.C layer slightly affects (increases) the calculated O/Nb+Sc ratio in the pre-exposed samples (Figure \ref{C4_methodology}a). The ambient stored samples are then vacuum annealed (Figure \ref{C4_methodology}b) in the processing chamber at 550 $^\circ$C for 2 h to saturate thermally induced processes before H*-exposures in the same chamber. The base pressure of the processing chamber is in the low 10$^{-8}$ mbar, while during annealing the maximum pressure of the chamber is in the low 10$^{-7}$ mbar range. The temperature of the sample is measured using an N-type thermocouple which is clamped on the sample surface. After annealing, the samples are cooled down to approximately 100 $^\circ$C and then transferred to the XPS chamber through a vacuum of low 10$^{-9}$ mbar. The XPS measurements on the annealed samples are referred to as pre-exposed (pre-exp) in the text and figures. The measurements on the pre-exposed samples are used as the reference for assessing the reduction of the samples upon H*-exposures.

The pre-exposed samples are then transferred through a vacuum back to the processing chamber for H*-exposure. H* in the chamber are generated through thermally cracking H$_2$ via W filament heated to $\approx$ 2000 $^\circ$C. For H*-exposure, the working pressure of the chamber is set to 0.02 mbar and the samples are placed $\approx$ 0.05 m from the W filament. The H* flux corresponding to these settings is calculated to be 10$^{21±1}$ H* m$^{-2}$s$^{-1}$ \cite{C4RN2233}. During H*-exposures, the sample temperature is maintained at 550 $^\circ$C. These H*-exposure conditions are relevant to fusion reactors and EUV scanners \cite{C4RN24, C4RN23, C4RN26, C4RN22}. 

Except for the Sc$_2$O$_3$ sample, the samples are exposed to H* for a total of 8 h, with XPS measurements taken after 2 h (Figure \ref{C4_methodology}c), 4 h (Figure \ref{C4_methodology}d), and 8 h (Figure \ref{C4_methodology}e). The Sc$_2$O$_3$ sample is exposed to H* for a total of 4 h, with XPS measurements taken after 10 min, 2 h, and 4 h. The XPS measurements on the H*-exposed samples are labeled according to the total H*-exposure time, i.e., as 10 min H*-exp, 2 h H*-exp, 4 h H*-exp, and 8 h H*-exp. Note that before each XPS measurement, the samples are cooled down to 100 $^\circ$C in the processing chamber and then transferred in-vacuum to the XPS chamber. 

The reducibility of the samples is evaluated based on the decrease in the ratio of at\% of O to at\% of Nb+Sc (O/Nb+Sc) upon H*-exposures. The O-, Nb-, and Sc-fractions in the samples are measured by effectively integrating the intensities (after background subtraction - calculated via the Shirley method) of their respective XPS spectra (O 1s, Nb 3d, and Sc 2p, respectively). These intensities are then scaled according to their respective Scofield sensitivity factors \cite{C4RNSF}. The decrease in the O/Nb+Sc ratios as a function of H*-exposure time, along with the changes in Nb 3d and Sc 2p XPS spectra taken at $\Theta$ = 34.25$^\circ$ are discussed in the text. Note that Nb 3d and Sc 2p XPS spectra presented in the main text figures are fitted with Voigt profile doublets, following Shirley background subtraction (the fitting method is detailed in the supplementary information). Furthermore, for better visualization, Nb 3d, and Sc 2p XPS spectra are normalized to the maximum intensity of Nb 3d and Sc 2p spectra of the pre-exposed samples, respectively.

Based on AR-XPS measurements, the chemical composition across the depth of the samples is found to be homogeneous. Variation in the O/Nb+Sc and Sc/Nb ratios as a function of $\Theta$ for each AR-XPS measurement is less than 10\% (Figure S4-S8). Therefore, we only discuss the XPS spectra taken at $\Theta = 34.25^\circ$ in the main text. The homogeneity of the pre-exposed NbScO$_4$ sample across the depth is further confirmed by cross-sectional Transmission Electron Microscopy (TEM) with Energy-Dispersive X-ray Spectroscopy (EDS) (Figure S2). The supplementary information further contains as-collected Nb 3d, Sc 2p, O 1s, and Si 2p XPS spectra as a function of $\Theta$ for each AR-XPS measurement (Figure S9-S13).

The work function of the samples is measured via XPS \cite{C4RN10}. To do this, we collect both the low kinetic energy (LKE) and valence band (VB) spectra at a negative bias of 16.4 V. This bias accelerates low kinetic energy (secondary) electrons from the sample towards the analyzer and separates these electrons from those scattering off the analyzer's wall. To assess whether the samples accumulate a net charge during measurements - particularly given the high dielectric constants of oxides - we also collect VB spectra of the samples without bias. The offset in the binding energies of the VB spectra collected with and without bias is almost equal to the applied bias. This suggests that our samples exhibit sufficient conductivity (Figure S14 and S15). Furthermore, LKE spectra exhibit a single secondary electron cutoff, suggesting that variation in the work function across the surface is insignificant (Figure S14) \cite{C4RN37}. Note that, in our setup, the electron analyzer and the sample's normal are not aligned, which introduces a systematic offset of $-$1 $\pm$ 0.2 eV in the measured work function \cite{C4RN10}. The work function values reported in the text and figures have already been adjusted to account for this offset. Furthermore, the uncertainty in measuring the secondary electron cutoff is $\pm$ 0.1 eV, resulting in an uncertainty of $\pm$ 0.2 eV in the difference between the measured work function values. 

To validate the work function measurements, we also measure the work function of the pre-exposed Nb$_2$O$_5$ and NbScO$_4$ samples using Kelvin Probe Atomic Force Microscopy (KPAFM). These work function measurements are in close agreement with those obtained using XPS (Figure S16). Furthermore, KPAFM measurements also suggest that the work function variation across the samples' surfaces is insignificant, consistent with the LKE XPS spectra.

\begin{acknowledgement}
This work has been carried out in the frame of the Industrial Partnership Program “X-tools,” Project No. 741.018.301, funded by the Netherlands Organization for Scientific Research, ASML, Carl Zeiss SMT, and Malvern Panalytical. We acknowledge the support of the Industrial Focus Group XUV Optics at the MESA+ Institute for Nanotechnology at the University of Twente. The authors extend their gratitude to Dr. Kai Sotthewes, Dr. Martina Tsvetanova, and Dr. Melissa J. Goodwin from the University of Twente for conducting the KPAFM and TEM-EDS measurements.
\end{acknowledgement}

\begin{suppinfo}
\begin{itemize}
  \item Supplementary information includes XRD of the pre-exposed samples, cross-sectional TEM of the pre-exposed NbScO$_4$ sample, KPAFM measurements of the pre-exposed Nb$_2$O$_5$ and NbScO$_4$ samples, calculated Sc/Nb and O/Nb+Sc ratios as a function of $\Theta$, fitting method for the Nb 3d and Sc 2p XPS spectra, comparison of the Nb 3d, Sc 2p, O 1s, and Si 2p XPS spectra as a function of $\Theta$, as well as LKE and VB spectra for the samples.
\end{itemize}

\end{suppinfo}

\bibliography{main}

\end{document}